# Extrinsic Rayleigh coefficient and domain mobility in critical MPB compositions of high-performance piezoceramics: A revisit

Mulualem Abebe and Rajeev Ranjan*

Department of Materials Engineering, Indian Institute of Science Bangalore-560012, India

**Abstract**

Enhanced domain mobility is considered as one of the important characteristics of morphotropic phase boundary (MPB) ferroelectrics exhibiting a very large piezoelectric response. For nearly two decades, Rayleigh analysis of dielectric and piezoelectric response at sub-coercive fields has been used as a phenomenological tool to determine the relative contribution of domain walls in influencing polar properties in these systems. The high value of the extrinsic Rayleigh coefficient ($\alpha$) at the MPB is generally attributed to enhanced domain mobility. Using coercive field as a measure of hindrance to domain mobility, in this paper, we have examined the correlation between the extrinsic Rayleigh coefficient $\alpha$, piezoelectric response, and coercive field ($E_C$) by a comparative study on three different high performance piezoelectrics− (BaTi$_{0.88}$Sn$_{0.12}$)-(Ba$_{0.7}$Ca$_{0.3}$)TiO$_3$, (ii) Pb(Zr,Ti)O$_3$ and (iii) PbTiO$_3$-BiScO$_3$ across their respective MPBs. Our study shows that although the three systems exhibit maximum piezoelectric response and maximum extrinsic Rayleigh coefficient $\alpha$ for their respective MPB compositions, the domain mobility in the MPB compositions of the Pb-based systems is not the highest. The lead-free system, on the other hand, exhibits highest domain mobility for the critical MPB composition. Contrary to the common perception, our study reveals that large value of the Rayleigh coefficient at the MPB is not necessarily indicative of enhanced domain mobility. We argue that the large piezoresponse in MPB systems is a result complex interplay of field induced motion of the interphase boundaries and domain walls. In the lead-based systems the interphase boundary motion hinders domain wall mobility. The large value of the Rayleigh coefficient and piezoresponse in these systems is because of the dominant role of the interphase boundary motion instead of the enhanced domain wall mobility.

*rajeev@materials.iisc.ernet.in

The toxicity associated with Pb in the commercial Pb(Zr, Ti)O$_3$ based piezoelectric has stimulated research on lead-free piezoelectrics [1-3]. A remarkable breakthrough was the discovery of very large piezoelectricity (d$_{33}$~620 pC/N) in (Ba,Ca)(Ti,Zr)O$_3$ ceramic by Liu and Ren in 2009 [3].  High piezo - response has also been reported in Sn and Hf modified BaTiO$_3$ systems [4-8]. The development of new high-performance lead-free piezoelectric also throws up challenges and an opportunity to study the underlying mechanisms therein, and compare it with the mechanisms reported before, mostly in the lead-based systems. Over the years different viewpoints have emerged to explain high piezo-response in MPB [9 - 22]. V. A. Isupov attributed the large piezoelectric response of the MPB to efficient poling enabled by the presence of many possible domain orientations in the two-phase state [10]. Other groups emphasized on domain wall density and their mobility [9, 11-13]. A different perspective was put forth by the polarization rotation model, according to which the large dielectric/piezoelectric response is due to rotation of the polarization vector along low energy pathways enabled by anisotropic flattening of the free energy profile [14 – 18]. In the framework of martensitic theory, enhanced contribution of domain walls arises due to a drastic decrease in domain wall energy at the MPB, leading to their increased density and mobility [19-.22]. The observation of nano-domains in MPB compositions have often been taken as proof of the validity of this concept [19, 20, 23, 24].

In the conventional terminology, the domain wall contribution is regarded as extrinsic, and the contribution due to atomic displacements within the unit cell is regarded as intrinsic [25]. The motion of the domain walls leads to hysteretic and non-linear dielectric and piezoelectric response at high fields [26 - 29]. For weak fields, domain wall motion is expected to be reversible. Field induced atomic displacements within the unit cell also contribute to the reversible component of the overall piezo-response. The macroscopic dielectric and piezoelectric response of a ferroelectric system due to motion of domain walls in a medium comprising of randomly distributed pinning centers of varying barrier heights can be described in terms of Rayleigh relationship [25, 30]. According to this relationship, for sub-coercive fields, the magnitude of the property is proportional to the amplitude of the field. The amplitude dependence of the converse piezoelectric coefficient is given by

$$d(E_o) = d_{rev} + \alpha_d E_o \qquad (1)$$

where $E_o$ is the amplitude of the cyclic sub-coercive electric field.  $d_{rev}$ and $\alpha E_o$ represent the reversible and the irreversible contributions, respectively. The validity of this relationship has been tested for a wide variety of piezoelectrics such as PZT [25, 29-33], $BaTiO_3$ [30], $PbTiO_3$-$BiScO_3$ [34], and $(Na,K)NbO_3$-based system [35].

Although not much discussion is available in literature, it is to be anticipated that additional complexity would come into picture with regard to the identification of the extrinsic and intrinsic contributing factors, if the system also undergoes stress/electric field induced phase transformation. Since the polarization direction is intimately related to the crystal structure of the ferroelectric phase, polarization rotation/switching accompanying free-energy flattening should be accompanied by structural transformation. With respect to the conventional terminology prevalent in literature, field/stress induced polarization rotation within a single domain would fall under the category of intrinsic response. In the recent years field induced phase transformations have been reported in several MPB systems [7, 36-40], which can be rationalized in the framework of polarization rotation/switching model. In real MPB systems, it is observed that one coexisting phase grows at the expense of the other on application of electric field. Evidently, this phenomenon would involve the movement of the interphase boundaries. In general, the interphase boundary motion is clubbed with the domain wall motion, and is regarded as contributing to the extrinsic response [18, 41]. In view of this, field induced structural transformation can be regarded as contributing to both the intrinsic and extrinsic response. The nature of the interaction between interphase boundary and domain walls has not received serious attention so far in the literature. Very recently, Lalitha *et. al*. have demonstrated that the domain switching propensity is rather reduced in the MPB composition of the piezoelectric system $(1-x)PbTiO_3$-$xBiScO_3$ in comparison to a non-MPB rhombohedral composition [42]. This observation correlates well with the lower coercive field reported for the non-MPB rhombohedral composition (Ec = 1.6 kV/mm) as compared to the critical MPB composition (1.9 kV/mm) [38]. Even for PZT, the data reported by Pramanick et al [33], suggest that although the piezoelectric response and the extrinsic Rayleigh parameter $\alpha_d$ exhibits maximum value for the MPB composition, the coercive field is still higher than that of the non-MPB rhombohedral composition. At the outset, these results seem to suggest that the large piezoelectric response of MPB compositions of PZT and PT-BS is not primarily associated with increased domain mobility. There is not enough study in this regard to establish if these seemingly contradictory results is true for other MPB systems or not. The issue needs to be further investigated in different MPB systems to help understand the detailed mechanism associated with large piezo-response. In this paper, we have carried out Rayleigh analysis of the converse piezoelectric response as a function of composition, close to the MPB of a high performance lead-free piezoelectric system $(1-x)(BaTi_{0.88}Sn_{0.12})$-$x(Ba_{0.7}Ca_{0.3})TiO_3$. This system was reported to exhibit very high piezoelectric response ($d_{33} \sim 530$ pC/N) for a composition exhibiting coexistence of tetragonal and rhombohedral phases [3]. We show that although, in conformity with PZT and PT-BS, the critical composition of BST-xBCT exhibits maximum extrinsic Rayleigh parameter ($\alpha_d$) and piezoelectric response, yet in contrast with the two lead-based systems, the coercive field of the critical MPB composition happens to be the lowest which seems to be in conformity with the predictions of martensitic theory based arguments. We have proposed plausible scenarios for the different behaviour in this lead-free, and the lead-based systems.

Different compositions of (1-x)(BaTi$_{0.88}$Sn$_{0.12}$)-x(Ba$_{0.7}$Ca$_{0.3}$)TiO$_3$ with x = 0.26, 0.27, 0.28, 0.29 and 0.30) were prepared by conventional solid-state reaction method. The starting materials were BaCO$_3$ (Alfa Aesar, with 99.8% purity), CaCO$_3$ (Alfa Aesar, with 99.95% purity), SnO$_2$ (Alfa Aesar, with 99.9% purity), and TiO$_2$ (Alfa Aesar, with 99.8% purity). The powders were mixed in stoichiometric ratio in acetone and ball milled for 12 h with an yttrium-stabilized zirconia ball as the milling medium. The mixed powders were calcined at 1100℃ for 4 h to form solid solutions. Thereafter, calcined powders were remixed using PVA and pressed into 13 mm diameter pellets and sintered at 1300℃ for 4 h and 1500℃ for 6 h. For electrical measurements, silver paste was coated on both side of the samples as the electrodes, and the samples were fired at 300 $^0$C for 1 h. All samples were poled at a constant poling time of 5 minutes and electric field amplitudes of 0.5kV/mm at room temperature in a silicon oil bath.

Fig. 1 (a, b) shows the polarization –field and strain-field hysteresis of BTS-xBCT ceramics at room temperature for different compositions. All the samples exhibit well defined hysteresis loops, which is characteristic of a ferroelectric material. The strain-electric-field ($S$-$E$) curves of the BST-xBCT ceramics show a typical ferroelectric butterfly-curve behavior. Fig. 1c shows the composition dependence of ferroelectric and piezoelectric properties of BTS-xBCT ceramics at room temperature. The relative permittivity, remnant polarization $P_r$, the saturation polarization $P_m$, the converse piezoelectric coefficient $d* = S_{max}/E_{max}$ at maximum measured field ($\sim$ 40 kV/mm, see Fig. 1b)) and the direct piezoelectric coefficient ($d_{33}$) measured by a Berlinocurt based parameter, all exhibits a maximum value of $x = 0.29$. The intrinsic and the extrinsic Rayleigh parameters were determined on all the compositions from the amplitude ($E_o$) dependence of the converse piezoelectric coefficient $d$ at subcoercive electric fields. $d$ was determined from the strain ($S$) versus electric field ($E$) hysteresis measurements at different sub-coercive field amplitudes ($E_o$) in the range 50 to 150V/mm, using $d = S_o/E_o$. Before the measurements, the pellets were first poled at a field of 500 V/mm for 5 minutes at room temperature and aged for 24 hours. Typical $S$-$E$ curves at different field amplitudes are shown in Fig. 2a for a representative composition x =0.29. *S-E* plots for the other compositions are shown in the supplemental file S1. The $d - E_o$ plots, thus derived for the four different compositions are shown in Fig. 2b. The linear nature of $d - E_o$ confirms the validity of the Rayleigh approach for this piezoelectric system. The intrinsic and the extrinsic Rayleigh parameters, $d_{rev}$ and $\alpha_d$, was calculated by fitting the Rayleigh relation given by equation (1). The composition dependence of both the parameters is shown in Figure 3(a). Both exhibit highest value for x = 0.29, the composition exhibiting the maximum piezoelectric response. This implies that both the intrinsic and the extrinsic contributions grows as this composition (x=0.29) is approached from either side. Similar to our case, Eitel et al [34] have reported a maximum value of $\alpha_d$ (3.3 x $10^{-16}$ m$^2$/V$^2$) for the MPB composition of a piezoelectric alloy PbTiO$_3$-BiScO$_3$ (PT-BS) exhibiting $d_{33}$ ~ 490 pC/N. The corresponding values for the tetragonal and the rhombohedral compositions outside of the MPB region was reported as 1.99 x $10^{-16}$ m$^2$/V$^2$ and 0.51 x $10^{-16}$ m$^2$/V$^2$, respectively [34]. A similar study by Pramanick et al [33] on 2 at % La-modified PZT shows $\alpha$ to be (in units of x $10^{-16}$ m$^2$/V$^2$) 4.3, 3.6 and 0.7 for the MPB (Zr:Ti :: 52:48), rhombohedral (Zr:Ti:: 40:60) and tetragonal (Zr:Ti::40:60), respectively. These compositions exhibit longitudinal piezoelectric coefficient $d_{33}$ as 480 pC/N, 208 pC/N and 130 pC/N, respectively. These independent studies clearly prove that the maximum $\alpha_d$ corresponds to the composition exhibiting the maximum piezoelectric response, and seems to be a universal feature. The piezoelectric coefficient and the extrinsic Rayleigh parameter $\alpha_d$ for PZT and PT-BS are tabulated as a function of composition, along with our results, in Table 1. We also calculated the ratio of the irreversible to the reversible Rayleigh parameter $\alpha_d/d_{rev}$ as a function of composition. This parameter determines the relative increase of $d$ per unit electric field due to irreversible displacement of the domain walls [30]. For our system this ratio exhibits a peak at x =0.29, suggesting that the largest irreversible contribution occurs in the most critical composition. A similar observation was made by Kim et al for PZT films close to the MPB [32].

In general, the increase in $\alpha$ is perceived as a manifestation of increase in the domain wall contribution. Based on Devonshire-Ginzburg –Landau theory for a polycrystalline ferroelectric ceramic Li et al [21] have argued that domain switching would be relatively easy for ferroelectric systems exhibiting multiple phase coexistence. In the framework of the martensitic theory, the enhancement in the piezoelectric response at the MPB is also attributed to the decrease in domain wall energy, resulting in their enhanced mobility [19, 20]. In ferroelectric measurements, coercive field ($E_C$) is a measure of hindrance to domain switching. If the increase in $\alpha_d$ is to be interpreted as an increase in domain wall mobility, one should expect $E_C$ to be minimum for the critical MPB composition. For our system, this seems to be the case, Fig. 3b. However, the same interpretation does not hold for PT-BS and PZT. Although Eitel et al have reported the largest $\alpha_d$ for the MPB composition of the PT-BS system [29], Lalitha et al have reported that the critical MPB composition x = 0.3725 of PT-BS, which exhibits a far superior piezoelectric response ($d_{33} \sim 425$ *pC/N*) shows $E_c = 1.9$ kV/mm. The rhombohedral non-MPB composition (x = 0.40), which exhibits a significantly lower $d_{33}$ (260 *pC/N*), shown a lower $E_C = 1.6$ kV/mm. [38]. Lalitha et al have recently proved using in-situ electric field dependent x-ray diffraction study that the domain switching propensity in the critical MPB composition the PT-BS system is significantly reduced as compared to the non-MPB rhombohedral/monoclinic composition [42], which seems to be consistent with the relatively higher *Ec* of the MPB composition of this system. A similar scenario was found while perusing the data for PZT, published by Pramanick et al (refer to Table I of ref. 33.). The 2 at % La modified non-MPB rhombohedral PZT (Zr/Ti:60/40) with $d_{33} = 208$ pC/N shows Ec =0.85 kV/mm. Whereas the MPB composition (Zr/Ti: 52: 48) with $d_{33} = 480$ *pC/N* shows a higher *Ec* of 1.35 kV/mm [33]. Thus, both the lead-based MPBs suggest relatively more hindrance to domain wall motion at the MPB as compared to the non-MPB rhombohedral composition. In contrast, for the BST-xBCT the most critical MPB composition shows the lowest hindrance to domain switching. These results suggest that there is no one-to-one correspondence between the large extrinisc Rayleigh parameter and the domain mobility (as measured in term of relative change in the coercive field) in different MPB systems.

Studies in the last few years have shown field induced phase transformation to be a common phenomenon of MPB ferroelectrics [36-39]. Kalyani et al have shown field induced tetragonal to orthorhombic transformation in Ba(Ti$_{0.98}$Sn$_{0.02}$)O$_3$ and associated the abrupt enhancement in the piezoelectric response at this composition to the enhance propensity of intrinsic polarization switching [7]. The authors have also shown this propensity to exist in pure BaTiO$_3$ and related the relatively large dielectric and piezoelectric properties of pure BaTiO$_3$ as compared to other known pure ferroelectric compounds such as KNbO$_3$, PbTiO$_3$ and Na$_{1/2}$Bi$_{1/2}$TiO$_3$ to this feature [43]. Field induced transformation and its role in the enhancement of piezoelectric response has also been demonstrated in the lead-free piezoelectric Na$_{1/2}$Bi$_{1/2}$TiO$_3$-BaTiO$_3$ [44]. However, unlike the BaTiO$_3$- based piezoelectrics, the structure-property correlation in Na$_{1/2}$Bi$_{1/2}$TiO$_3$ based systems has an added complexity due to the presence of intrinsic structural inhomogeneity on the mesoscopic scale in the parent compound, and its significant role in determining the global structures and macroscopic polar properties [45-48]. In view of the common occurrence of the field induced phase transformations in different MPB systems, this phenomenon is also expected in the critical composition of $(1-x)(BaTi_{0.88}Sn_{0.12})-x(Ba_{0.7}Ca_{0.3})TiO_3$. Xue et al have reported that the critical MPB composition of this system exhibits a coexistence of tetragonal and rhombohedral phases [4]. The electric field will make one of the coexisting phases grow at the expense of the other. For PZT, it has been reported that the fraction of the rhombohedral/monoclinic phase increases with field [36, 37], whereas for PT-BS, the rhombohedral/monoclinic fraction decreases with field [38, 42]. The reduced domain mobility in the MPB compositions of PT-BS and PZT, as compared to the rhombohedral non-MPB phase, can be associated to the hindrance provided to the domain mobility by interphase boundary motion associated with the field induced phase transformations. The large value of the extrinsic Rayleigh parameter $\alpha$ of the MPB compositions of the two Pb-based systems must therefore be attributed primarily to the interphase boundary motion. However, the fact that the lowest coercive field is obtained for the most critical MPB composition in the lead-free BST-xBCT suggests that the interphase boundary motion, if there is any, as a result of field induced transformation, does not necessarily impede domain mobility in this system. In view of this, it would be interesting to study the relative change in the domain switching fraction in the coexisting tetragonal and rhombohedral phases, as the critical composition is approached by field dependent in-situ high energy x-ray diffraction. Our study suggests that there is a competition between domain wall and interphase boundary motions, and depending on the system, the interphase boundary motion may or may not hinder the domain mobility. We hope that this work will stimulate further studies for more detailed investigation.

In conclusion, a combined analysis of piezoelectric response, extrinsic Raleigh parameter, and coercive field was carried out on three different high performance piezoelectric systems, PZT, $PbTiO_3-BiScO_3$ and $(BaTi_{0.88}Sn_{0.12})-(Ba_{0.7}Ca_{0.3})TiO_3$. All the systems show largest extrinsic Rayleigh parameter ($\alpha$) for the composition exhibiting the largest piezo-response. Using coercive-field as a measure of hindrance to domain motion, we found that for the lead-free system domain mobility is maximum at the critical MPB composition, as anticipated in the martensitic theory based explanation of high piezoelectric response. In contrast, for the Pb-based systems, the domain mobility is lower in the MPB composition, as compared to the non-MPB rhombohedral composition. This study, therefore, rules out a universal correlation between extrinsic Rayleigh parameter ($\alpha_d$), and domain mobility in high performance piezoelectrics. It is argued that the large $\alpha_d$, in the MPB compositions of the Pb-based systems is a manifestation of dominant role of the field induced interphase boundary motion. Our study suggests that interphase boundary motion hinders domain mobility in the Pb-based systems, whereas this hindrance is not as effective in the lead-free system.

**Acknowledgment:** RR acknowledges Science and Engineering Board (SERB) of the Department of Science and Technology, Govt. of India (SERB/F/5046/2013-14) and ISRO-IISc, Space Technology Cell for financial assistance.

Table1. Piezoelectric coefficient $d_{33}$ and the Rayleigh parameter α and coercive field ($E_C$) in different MPB systems as a function of composition.

| System | $\alpha$ (x10$^{-16}$ m$^2$/V$^2$) | $d_{33}$ (pC /N) | $E_C$ (kV/ mm) | Reference |
|---|---|---|---|---|
| PLZT5248 (MPB) | 4.3 | 480 | 1.35 | Pramanick *et. al.*, ref [33] |
| PLZT6040 (Rombohedr) | 3.6 | 208 | 0.85 | |
| PLZT4060 (Tetragonal) | 0.70 | 130 | 2.60 | |
| BS-0.62PT (Rhomb.) | 1.74 | 272 | | Eitel *et. al.*, ref [34] |
| BS-0.64PT (MPB) | 3.34 | 490 | | |
| BS-0.70PT (Tetragonal) | 0.82 | 165 | | |
| PT-xBS x =0.40 (Rhomb.) | | 260 | 1.6 | Lalitha *et. al.* Ref. [38] |
| PT-xBS x =0.3725 (MPB) | | 425 | 1.9 | |
| BTS-0.26BCT | 6.3 | 352 | 2.73 | Present work |
| BTS-0.27BCT | 7.5 | 475 | 2.6 | |
| BTS-0.28BCT | 13.2 | 485 | 2.5 | |
| BTS-0.29BCT | 16.6 | 511 | 2.3 | |
| BTS-0.30BCT | 9.8 | 470 | 2.5 | |

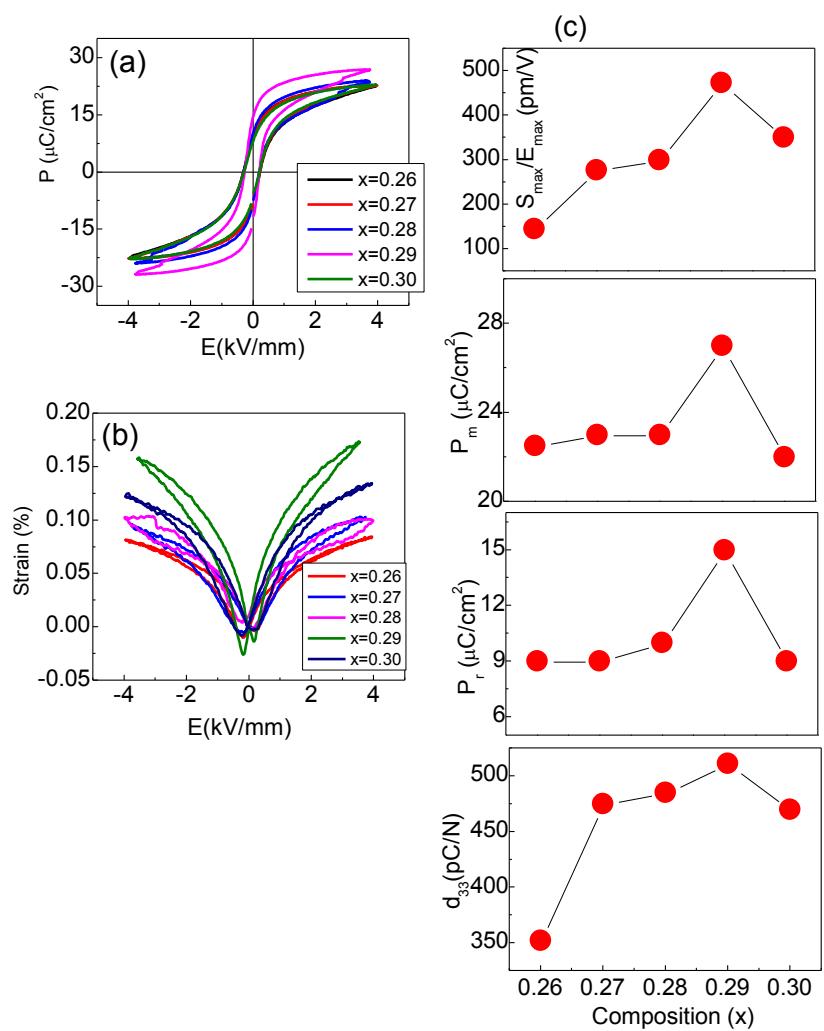

Figure 1 (a) shows the polarization −field and (b) strain-field hysteresis of BTS-xBCT ceramics at room temperature for different compositions (x). Fig. 1c shows the composition dependence of ferroelectric and piezoelectric properties of BTS-xBCT ceramics at room temperature. $d_{33}$ is direct piezoelectric coefficient as measured by Berlincourt meter; $P_r$ is remanent polarization and $Pm$ is maximum polarization.

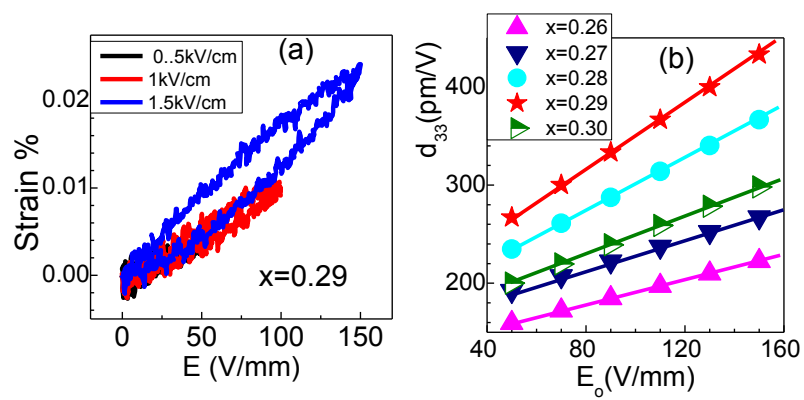

Figure 2 (a) shows a typical *S-E* curves at different field amplitudes for a representative composition x =0.29. (b) shows the converse piezoelectric coefficient (*d*) of BTS-xBCT ceramic  as a function of composition and electric field, calculated from the *S-E* data using the equation $d = S_o/E_o$, where $E_o$ is the amplitude of the field and $S_o$ is the strain at $E_o$

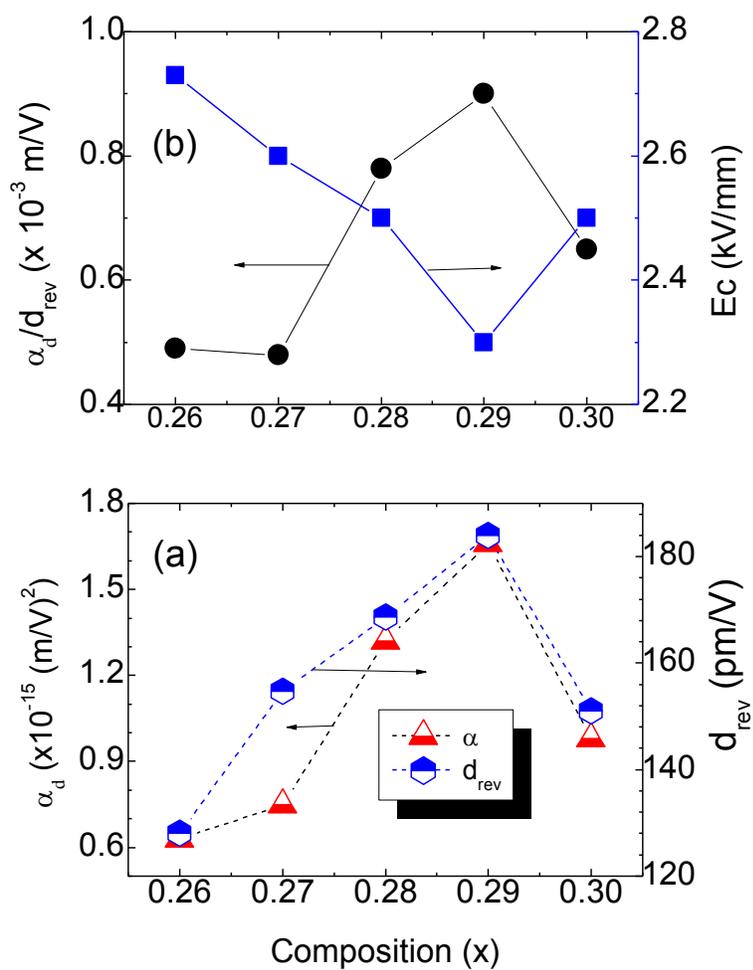

Fig. 3 (a) shows composition dependence of $\alpha$ and $d_{rev}$ of BTS-xBCT. (b) shows the ratio of irreversible to the reversible Rayleigh coefficient and coercive field as a function of composition.

**Supplemental information**

S1: Strain versus electric field at different sub-coercive field amplitudes in different compositions of (1-x) (BaTi$_{0.88}$Sn$_{0.12}$)-x(Ba$_{0.7}$Ca$_{0.3}$)TiO$_3$ (BST-xBCT). The piezoelectric strain coefficient obtained from these measurements are shown in the adjacent plot on the right side of the corresponding S-E curves. Rayleigh parameters d$_{rec}$ (in unit of pm/V) and $\alpha_d$ (in unit of 10$^{-15}$ pm$^2$/V$^2$) were obtained by fitting Rayleigh relation to the $d - E_o$ linear plots.

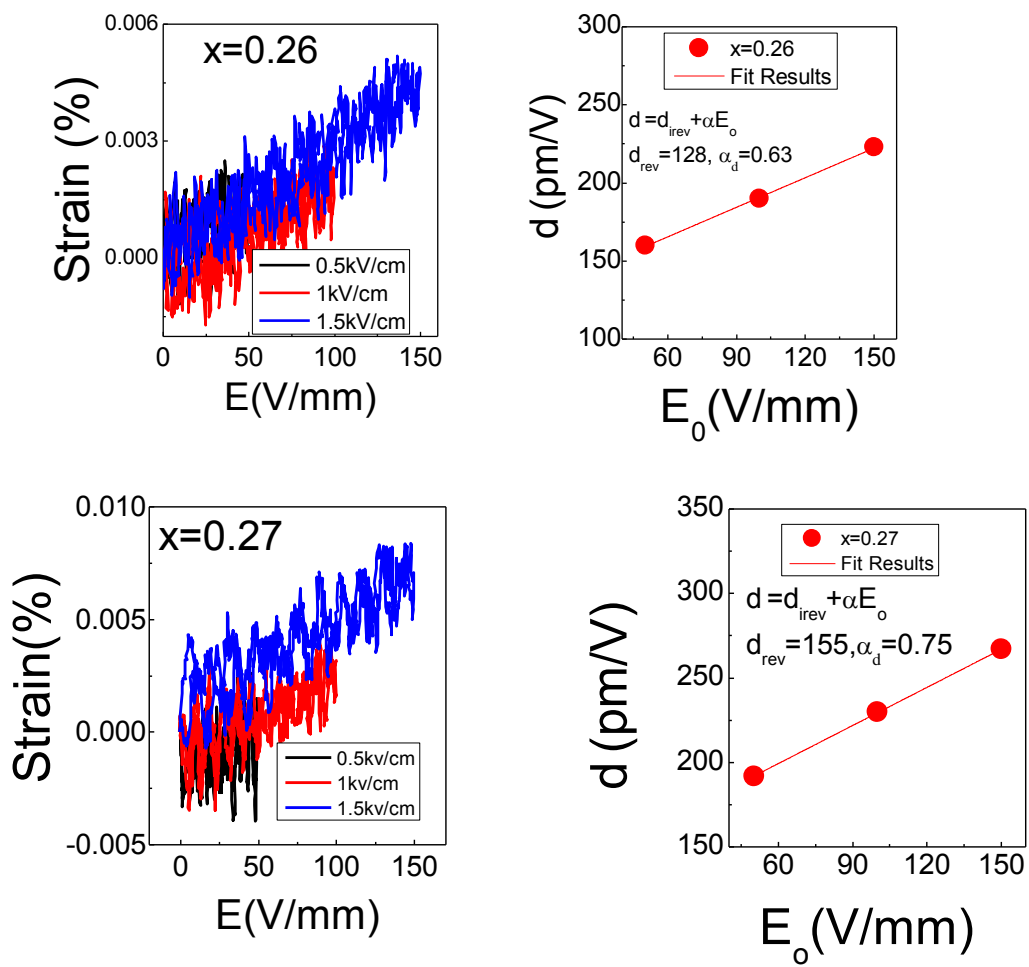

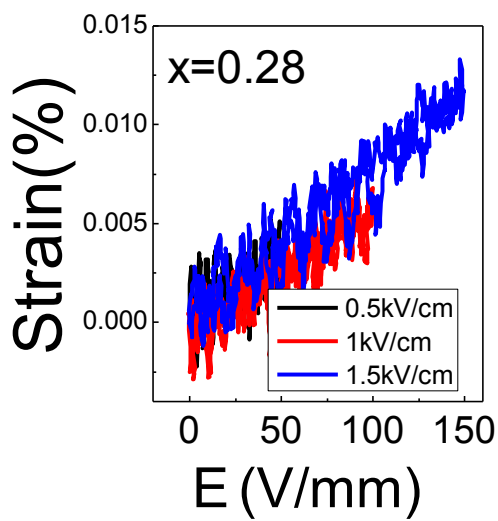

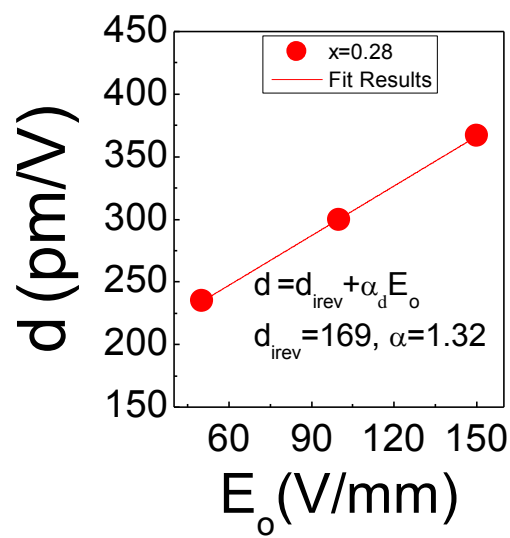

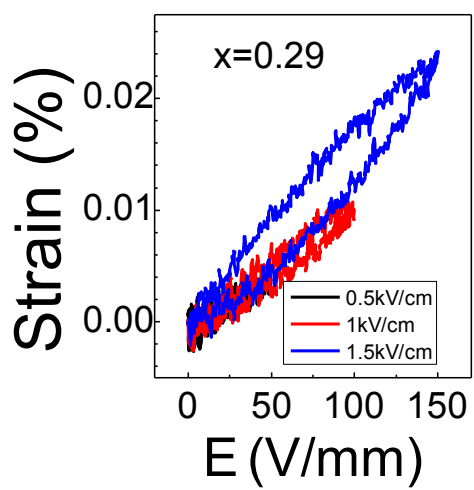

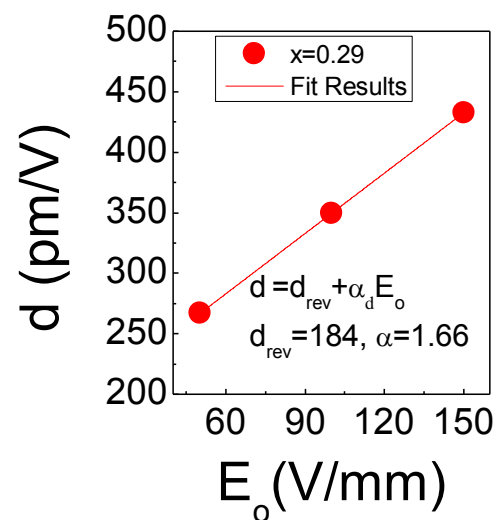

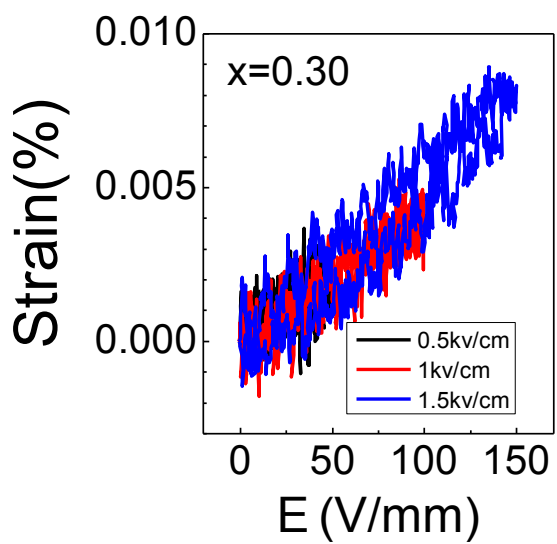

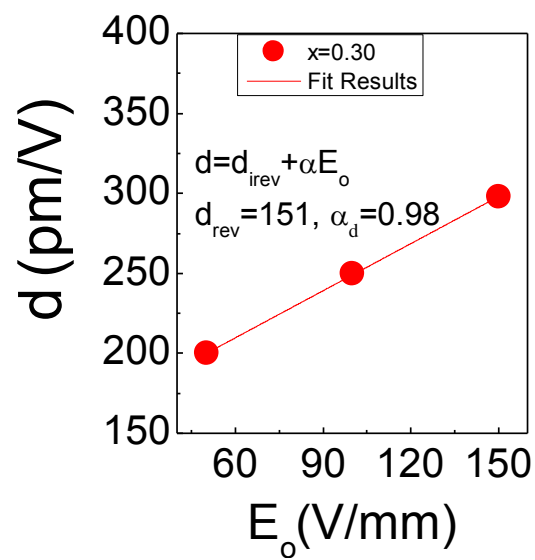